\documentclass[a4paper,11pt]{article}
\usepackage{pos}

\title{Transverse momentum-dependent parton distributions for
longitudinally polarized nucleons from domain wall fermion
calculations at the physical pion mass}
\ShortTitle{TMDs for longitudinally polarized nucleons 
at the physical pion mass}

\author*[a]{M.~Engelhardt}
\author[b]{N.~Hasan}
\author[c]{T.~Izubuchi}
\author[d]{C.~Kallidonis}
\author[e,f,g]{S.~Krieg}
\author[h]{S.~Meinel}
\author[i]{J.~Negele}
\author[i]{A.~Pochinsky}
\author[b,j]{G.~Silvi}
\author[k]{S.~Syritsyn}

\affiliation[a]{Department of Physics, New Mexico State University,
Las Cruces, NM 88003, USA}
\affiliation[b]{Bergische Universit\"at Wuppertal, 42119 Wuppertal, Germany}
\affiliation[c]{Physics Department, Brookhaven National Laboratory,
Upton, NY 11973, USA}
\affiliation[d]{Thomas Jefferson National Accelerator Facility,
Newport News, VA 23606, USA}
\affiliation[e]{JARA \&\, IAS, J\"ulich Supercomputing Centre,
Forschungszentrum J\"ulich, 52425 J\"ulich, Germany}
\affiliation[f]{Helmholtz-Institut f\"ur Strahlen- und Kernphysik,
Universit\"at Bonn, 53115 Bonn, Germany}
\affiliation[g]{Center for Advanced Simulation and Analytics (CASA),
Forschungszentrum J\"ulich, 52425 J\"ulich, Germany}
\affiliation[h]{Department of Physics, University of Arizona, Tucson,
AZ 85721, USA}
\affiliation[i]{Center for Theoretical Physics, Massachusetts Institute of
Technology, Cambridge, MA 02139, 
USA}
\affiliation[j]{J\"ulich Supercomputing Centre,
Forschungszentrum J\"ulich, 52425 J\"ulich, Germany}
\affiliation[k]{Department of Physics and Astronomy, Stony Brook University,
Stony Brook, NY 11794, USA}

\emailAdd{engel@nmsu.edu}

\abstract{Previous Lattice QCD calculations of nucleon transverse
momentum-dependent parton distributions (TMDs) focused
on the case of transversely polarized nucleons, and thus
did not encompass two leading-twist TMDs associated with
longitudinal polarization, namely, the helicity TMD and
the worm-gear TMD corresponding to transversely polarized
quarks in a longitudinally polarized nucleon. Based on a
definition of TMDs via hadronic matrix elements of quark
bilocal operators containing staple-shaped gauge connections,
TMD observables characterizing the aforementioned two TMDs
are evaluated, utilizing an RBC/UKQCD domain wall fermion
ensemble at the physical pion mass.}

\FullConference{%
  The 39th International Symposium on Lattice Field Theory (Lattice2022),\\
  8-13 August, 2022 \\
  Bonn, Germany 
}

\begin{document}
\maketitle

\section{Introduction}
Transverse momentum-dependent parton distribution functions \cite{revtmd}
(TMDs) embody fundamental information on the three-dimensional partonic
structure of hadrons. Complementary to generalized parton distributions
(GPDs), which provide insight into the transverse spatial structure
through their Fourier transforms, impact parameter-dependent parton
distributions, TMDs encode the distribution of transverse momentum among
the partons. From the point of view of a Lorentz frame in which a
nucleon of mass $m_N $ is propagating in the 3-direction with a large
momentum, $P^{+} \equiv (P^0 + P^3 )/\sqrt{2} \gg m_N $, the minus component
of the quark momentum $k$ becomes ignorable,
$k^{-} \equiv (k^0 - k^3 )/\sqrt{2} \ll m_N $, as a consequence
of which TMDs principally depend on the parton longitudinal momentum
fraction $x=k^{+} /P^{+} $ and the parton transverse momentum vector
$k_T $. Generic TMDs $f(x,k_T )$ are regarded as being integrated
over the $k^{-} $ component.

The hadronic structure encoded in TMDs inherently influences angular
asymmetries observed in experimental processes such as semi-inclusive
deep inelastic scattering (SIDIS) and the Drell-Yan (DY) process.
Motivated by past observations at COMPASS, HERMES and JLab
\cite{compass,hermes,clas}, current and upcoming experimental efforts
within the JLab 12 GeV program and the planned electron-ion collider
(EIC) place a strong emphasis on the further investigation of TMDs.
Central to the analysis of the experimental signatures and the connection
to the underlying hadronic structure distilled into the TMDs is a
factorization framework. A prominent such framework is the one developed
in \cite{spl1,collbook,spl2,same}, which is simultaneously well suited to
connect the QCD definition of TMDs to Lattice QCD calculations, as
will be laid out in more detail below. A feature of TMD factorization
that has attracted considerable interest, and that illustrates its
enhanced complexity compared to standard collinear factorization, is the
process dependence of TMDs introduced by initial or final state
interactions between the active quark parton and the hadron remnant. In
particular, the relative sign change of the Sivers TMD $f_{1T}^{\perp } $,
as extracted from SIDIS vs.~DY in transversely polarized nucleons, has
received intense scrutiny \cite{sivchg1,sivchg2,sivchg3}.

Not least as a consequence of the focus on this effect and others
observed in transversely polarized nucleons, also past lattice TMD
calculations \cite{tmdlat,bmlat,rentmd} have concentrated on the case of
transverse nucleon polarization. This has left the longitudinally polarized
sector largely unexplored, with the notable exception of generalized
TMD (GTMD) investigations of quark orbital angular momentum in the nucleon
\cite{f14pap,f14deriv,ls}. The present study closes this gap; all
leading-twist quark TMD observables have thereby now been explored
within the lattice TMD program previously developed in
\cite{tmdlat,bmlat,rentmd}.

\section{Construction of TMD observables}
\begin{figure}
\begin{center}
\includegraphics[width=7cm]{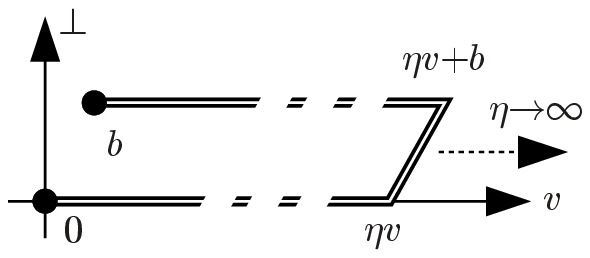}
\end{center}
\caption{Path of the Wilson line ${\cal U} $ in the matrix element
(\ref{spacecorr}).}
\label{figstaple}
\end{figure}

A construction of TMD ratio observables that to a large extent avoids
the complications that arise in the definition of isolated TMDs
\cite{lamettmd1,lamettmd2,lamettmd3}, and is consequently comparatively
straightforward to implement within Lattice QCD, is described in detail
in \cite{tmdlat}. To summarize, TMDs are defined based on the fundamental
nonlocal correlator
\begin{equation}
\widetilde{\Phi }^{[\Gamma ]}_{\mbox{\scriptsize unsubtr.} } (b,P,S,\ldots )
\equiv \frac{1}{2} \langle P,S | \ \bar{q} (0) \
\Gamma \ {\cal U} [0,\eta v, \eta v+b,b] \ q(b) \ |P,S\rangle
\label{spacecorr}
\end{equation}
evaluated in a hadron state with momentum $P$ and spin $S$. The quark
operator separation $b$ is Fourier conjugate to the quark momentum $k$.
$\Gamma $ denotes an arbitrary Dirac $\gamma $-matrix structure, and
the quark operators are connected by a staple-shaped Wilson line ${\cal U}$
composed of straight-line segments running between the positions specified
in its argument, as also illustrated in Fig.~\ref{figstaple}. The direction
of the staple legs is given by the unit vector $v$, whereas their length is
parametrized by $\eta $ (below, both positive and negative values of $\eta $
will be allowed in order to encode a reversal of direction of the staple legs
in a simple manner). An important complication in the definition of TMDs
arises owing to the divergences associated with ${\cal U}$, which must
be compensated by a ``soft factor''. This soft factor is multiplicative and
can thus be cancelled, along with additional wave function renormalization
factors associated with the quark operators, by forming appropriate TMD
ratios. This is the scheme pursued in the following, as a result of which
specific consideration of the soft factor is avoided.

Furthermore, the most naive choice of $v$ as a light-cone vector introduces
rapidity divergences, which in the scheme advanced in
\cite{spl1,collbook,spl2} are regulated by taking $v$ off the light cone
into the space-like region, the approach to the light cone being governed
by perturbative evolution equations. In the following, the Collins-Soper
type evolution parameter measuring the deviation of $v$ from the light cone
will be chosen as
\begin{equation}
\hat{\zeta } = \frac{v \cdot P}{|v|\, |P|} \ .
\end{equation}
The light-cone limit corresponds to $\hat{\zeta } \rightarrow \infty $.
This prescription of employing space-like $v$ renders the connection to
Lattice QCD particularly straightforward, as detailed below.

A crucial further element of the present treatment is the decomposition of
the correlator $\widetilde{\Phi }^{[\Gamma ]}_{\mbox{\scriptsize unsubtr.} } $
defined in (\ref{spacecorr}) into Lorentz-invariant amplitudes, denoted
below by $\widetilde{A}_{iB} $. This decomposition will allow for a simple
translation of results between the Lorentz frame in which TMDs are
defined, as discussed up to this point, and the Lorentz frame suited
for a Lattice QCD calculation, introduced further below. At leading twist,
the decomposition for nucleons reads \cite{tmdlat}
\begin{eqnarray}
\frac{1}{2P^{+} }
\widetilde{\Phi }^{[\gamma^{+} ]}_{\mbox{\scriptsize unsubtr.} } &=&
\widetilde{A}_{2B} + im_N \epsilon_{ij} b_i S_j \widetilde{A}_{12B}
\nonumber \\
\frac{1}{2P^{+} }
\widetilde{\Phi }^{[\gamma^{+} \gamma^{5} ]}_{\mbox{\scriptsize unsubtr.} }
&=& -\Lambda \widetilde{A}_{6B} +i[(b\cdot P) \Lambda - m_N (b_T \cdot S_T )]
\widetilde{A}_{7B}
\label{adecomp2} \\
\frac{1}{2P^{+} }
\widetilde{\Phi }^{[i\sigma^{i+} \gamma^{5} ]}_{\mbox{\scriptsize unsubtr.} }
&=& im_N \epsilon_{ij} b_j \widetilde{A}_{4B} -S_i \widetilde{A}_{9B}
-im_N \Lambda b_i \widetilde{A}_{10B}
+ m_N [(b\cdot P) \Lambda - m_N (b_T \cdot S_T )] b_i \widetilde{A}_{11B} \ ,
\nonumber
\end{eqnarray}
where $\Lambda $ denotes the nucleon helicity (i.e.,
$S^{+} =\Lambda P^{+} /m_N $, $S^{-} =-\Lambda m_N /2P^{+} $).
Specializing to the case of longitudinal polarization, $S_T =0$, and
focusing on momentum fraction $x$-integrated TMDs, accessed in
$b$-space by considering $b\cdot P=0$, the amplitudes
$\widetilde{A}_{2B} $, $\widetilde{A}_{6B} $ and $\widetilde{A}_{10B} $
can be readily isolated\footnote{The amplitude $\widetilde{A}_{4B} $
associated with the Boer-Mulders effect, observed in unpolarized hadrons,
will not be considered further here, since it has already been discussed
in previous studies focusing on transversely polarized or unpolarized
hadrons \cite{tmdlat,bmlat,rentmd}.} and serve to define the following
TMD ratio observables in which multiplicative soft factors associated
with the Wilson line ${\cal U}$ and wave function renormalization factors
attached to the quark operators in (\ref{spacecorr}) at finite separation
$b$ are cancelled:
\begin{itemize}
\item
The generalized $h_{1L}^{\perp } $ worm-gear shift
\begin{equation}
\langle k_x \rangle_{LT} (b_T^2 , \ldots ) =
-m_N \widetilde{A}_{10B} (-b_T^2 , \ldots ) /
\widetilde{A}_{2B} (-b_T^2 , \ldots ) =
m_N \tilde{h}_{1L}^{\perp [1](1)} (b_T^2 , \ldots ) /
\tilde{f}_{1}^{[1](0)} (b_T^2 , \ldots )
\label{wgshift}
\end{equation}
where the expression on the right-hand side establishes the connection
to Fourier-transformed TMD moments, cf.~\cite{tmdlat}. In the formal
$b_T \rightarrow 0$ limit, (\ref{wgshift}) represents the average
transverse momentum $k_x $ of quarks polarized in the same transverse
(``$T$'') direction, i.e., here, the $x$-direction, in a longitudinally
(``$L$'') polarized nucleon, normalized to the number of valence quarks.
It should be noted that the $b_T^2 =0$ limit introduces additional
divergences that require adequate treatment compared to the (``generalized'')
finite $b_T $ case.
\item
The generalized axial charge
\begin{equation}
\Delta \Sigma (b_T^2 ,\ldots ) =
-\widetilde{A}_{6B} (b_T^2 ,\ldots ) /
\widetilde{A}_{2B} (b_T^2 ,\ldots ) =
\tilde{g}_{1}^{[1](0)} (b_T^2 ,\ldots ) /
\tilde{f}_{1}^{[1](0)} (b_T^2 ,\ldots )
\label{genac}
\end{equation}
which, in the $b_T \rightarrow 0$ limit, reduces to the standard
axial charge, normalized to the number of valence quarks. In this case,
the additional divergences in the $b_T^2 =0$ limit are associated with
the standard renormalization factors $Z_A $ and $Z_V $ in the numerator
and denominator, respectively; using domain wall fermions, as is done
in the following lattice calculation, one has $Z_A = Z_V $ and these
renormalization factors cancel\footnote{Preserving chiral symmetry in
the fermion discretization moreover prevents the appearance of operator
mixing effects \cite{constmixtmd,shanamixtmd,greenmixtmd,jimixtmd}
induced by the breaking of chiral symmetry. Such effects would
invalidate the cancellation of renormalization factors through taking
ratios such as (\ref{wgshift}) and (\ref{genac}), since they would
generate additional additive terms in the numerators and denominators.}.
\end{itemize}

\section{Lattice calculation}
In a Lattice QCD calculation, the operator of which one is taking a
hadronic matrix element in (\ref{spacecorr}) has to be located at one
single Euclidean insertion time. On the other hand, the operator as
defined above in general extends into the (Minkowski) temporal
direction. However, since all separations in the operator, controlled
by the four-vectors $b$ and $v$, are space-like, the problem can be
boosted to a Lorentz frame in which the entire operator indeed exists
at a single time, and the lattice calculation can be performed in that
frame. This is the point where a formulation employing a space-like
direction $v$ for the purpose of regulating rapidity divergences is
crucial. Furthermore, having decomposed the resulting
$\widetilde{\Phi }^{[\Gamma ]}_{\mbox{\scriptsize unsubtr.} } $
into the invariant amplitudes $\widetilde{A}_{iB} $, the results
for those amplitudes are immediately also applicable in the
Lorentz frame in which (\ref{spacecorr}) was originally defined;
TMD observables of the type (\ref{wgshift}) and (\ref{genac})
are thus determined.

The data obtained in a lattice calculation are necessarily limited
to a finite staple length $\eta $ and a finite rapidity parameter
$\hat{\zeta } $, and they must be extrapolated to the limits
$\eta \rightarrow \infty $ and $\hat{\zeta } \rightarrow \infty $.
As will be seen below, the former extrapolation can be controlled
for a substantial range of the parameter space employed in this
work. The latter extrapolation is more challenging, since
large $\hat{\zeta } $ can only be achieved using large hadron
momentum $P$, and the set of $P$ that can be accessed reliably in a
lattice calculation is fairly limited. In the present study, no
quantitative extrapolation will be attempted, but data will be presented
for two values of $\hat{\zeta } $ that suggest that the variation with
$\hat{\zeta } $ is mild. Quantitative extrapolations were presented
in the dedicated study \cite{bmlat} of the Boer-Mulders effect in the
pion.

The present investigation utilized a 2+1 flavor domain wall fermion
ensemble provided by the RBC/UKQCD Collaboration, on a $48^3 \times 96$
lattice with spacing $a=0.114\, \mbox{fm}$, with pion mass close to the
physical limit, $m_{\pi} = 139\, \mbox{MeV} $ \cite{rbcukqcd}. On the 130
gauge configurations in the ensemble, an all-mode averaging scheme \cite{ama}
was implemented, evaluating 33280 low-accuracy and 520 exact samples for
bias correction. A significant source of systematic bias, evident in some
of the results discussed below, is the rather small source-sink separation
$8a=0.91\, \mbox{fm} $ that was employed in this exploratory calculation to
control statistical fluctuations. Two nucleon momenta, $P_3 = 2\pi /L$
and $P_3 = 4\pi /L$ (where $L=48a$ denotes the spatial extent of the
lattice), were used, corresponding to $\hat{\zeta } =0.23$ and
$\hat{\zeta } =0.46$.

\begin{figure}
\begin{center}
\includegraphics[width=7.3cm]{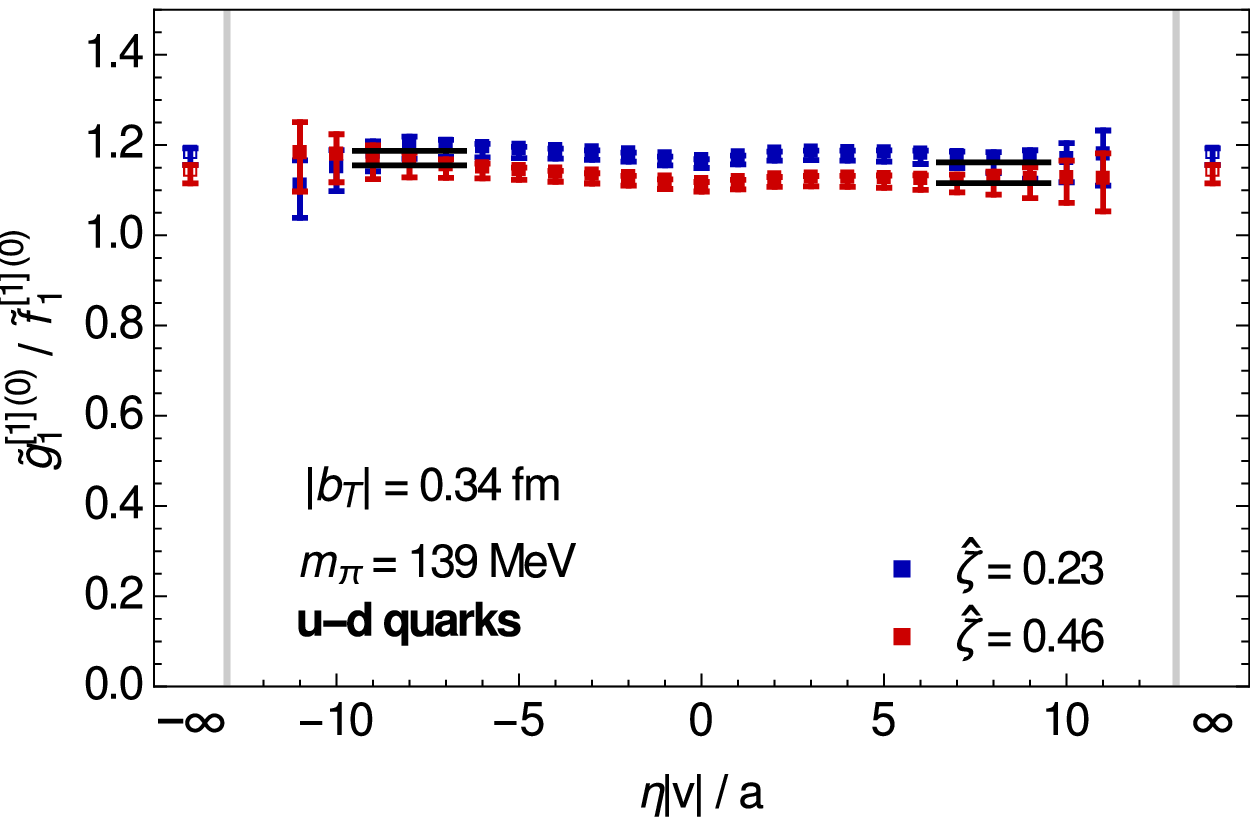}
\hspace{0.3cm}
\includegraphics[width=7.3cm]{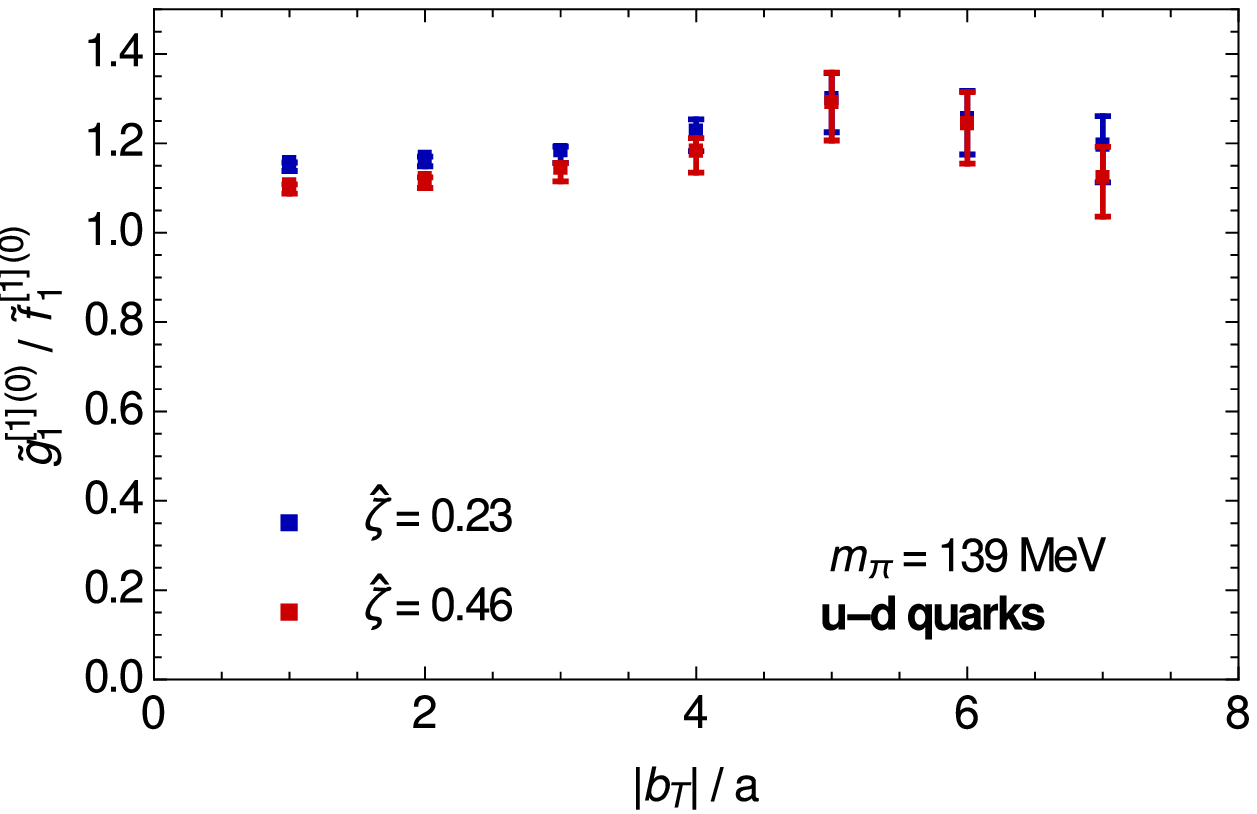}
\end{center}
\caption{Isovector generalized axial charge in the nucleon. Left: Dependence
on the staple length $\eta $ for fixed $|b_T |$ and two values of
$\hat{\zeta } $; the extrapolation to the $|\eta | \rightarrow \infty $
SIDIS/DY limit is described in the main text. Right: SIDIS/DY limit
as a function of $|b_T |$ for two values of $\hat{\zeta } $. Error bars
show statistical uncertainties only.}
\label{figetag}
\end{figure}

Fig.~\ref{figetag} shows results for the isovector generalized axial charge
in the nucleon. In the isovector, $u-d$ quark combination, diagrams in which
the operator is inserted into a disconnected quark loop cancel; such
diagrams were not evaluated in this study. The left-hand panel displays the
dependence on the staple length $\eta $ for a fixed operator separation
$|b_T |$; data for two values of $\hat{\zeta } $ are shown. Evidently, the
$\eta $-dependences of the moments of the unpolarized and helicity TMDs
$f_1 $ and $g_1 $ closely track one another, leading to virtually constant
behavior of the ratio. Extrapolation to $|\eta | \rightarrow \infty $ is
achieved by the plateau fits indicated by the horizontal lines, which are
averaged over the plateaus at positive and negative $\eta $ in view of the
T-even nature of the observable. $\eta \rightarrow \infty $ corresponds to
the SIDIS limit, whereas $\eta \rightarrow -\infty $ corresponds to the DY
limit. The data at the two values of $\hat{\zeta } $ do not differ to a
statistically significant extent, suggesting a merely mild dependence on
$\hat{\zeta } $. The right-hand panel summarizes the
$|\eta | \rightarrow \infty $ SIDIS/DY limits as a function of $|b_T |$.
Also the dependence on $|b_T |$ is mild and not distinguishable from a
constant to statistically significant degree. However, the magnitude of
the generalized axial charge obtained here betrays a systematic bias
that is presumably owed to excited state contaminations resulting from
the small source-sink separation used in the calculation. The data suggest
a $b_T =0 $ limit significantly below the experimental value
$\Delta \Sigma (b_T =0) = 1.272$. Such deficits have been widely
reported and analyzed in the literature \cite{ga1,ga2,ga3}
and thus their appearance in the present calculation does not come as a
surprise. Considerable, dedicated efforts are ultimately required to
resolve these discrepancies (see, e.g., \cite{ga3}).

\begin{figure}
\begin{center}
\includegraphics[width=7.3cm]{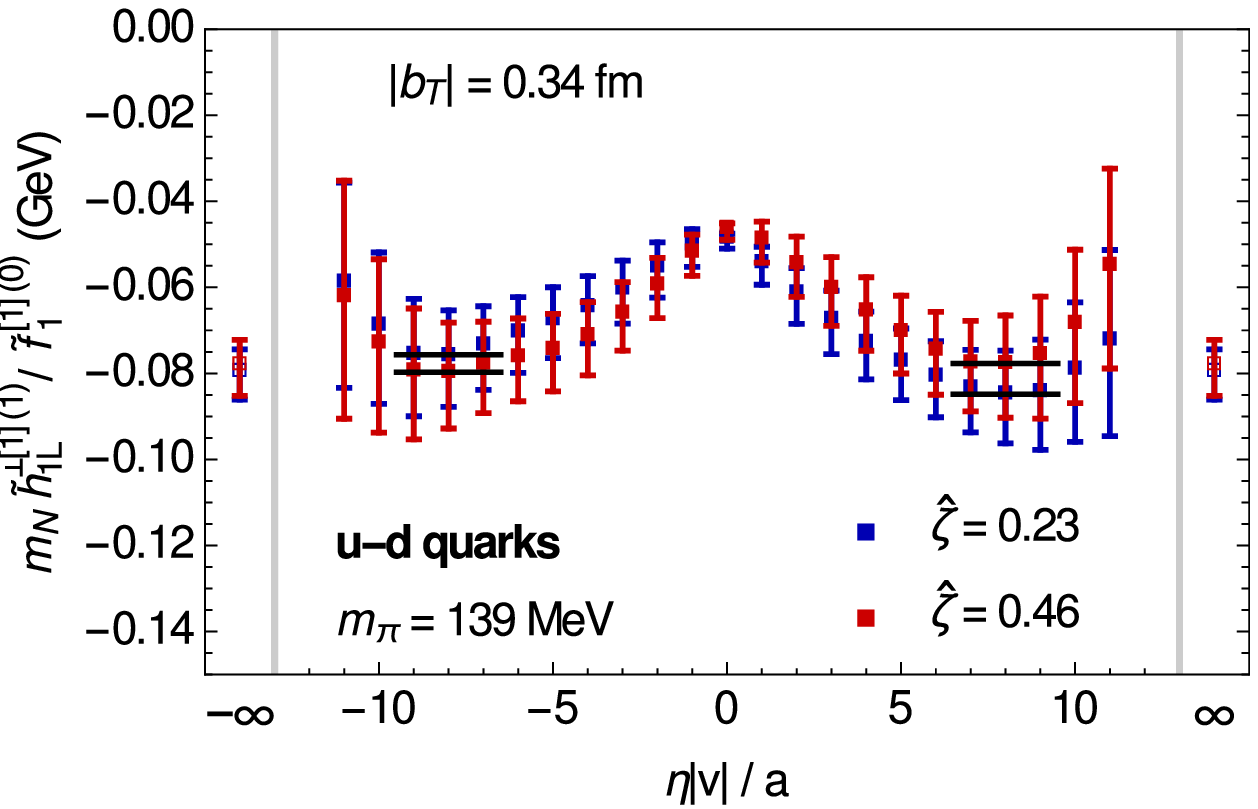}
\hspace{0.3cm}
\includegraphics[width=7.3cm]{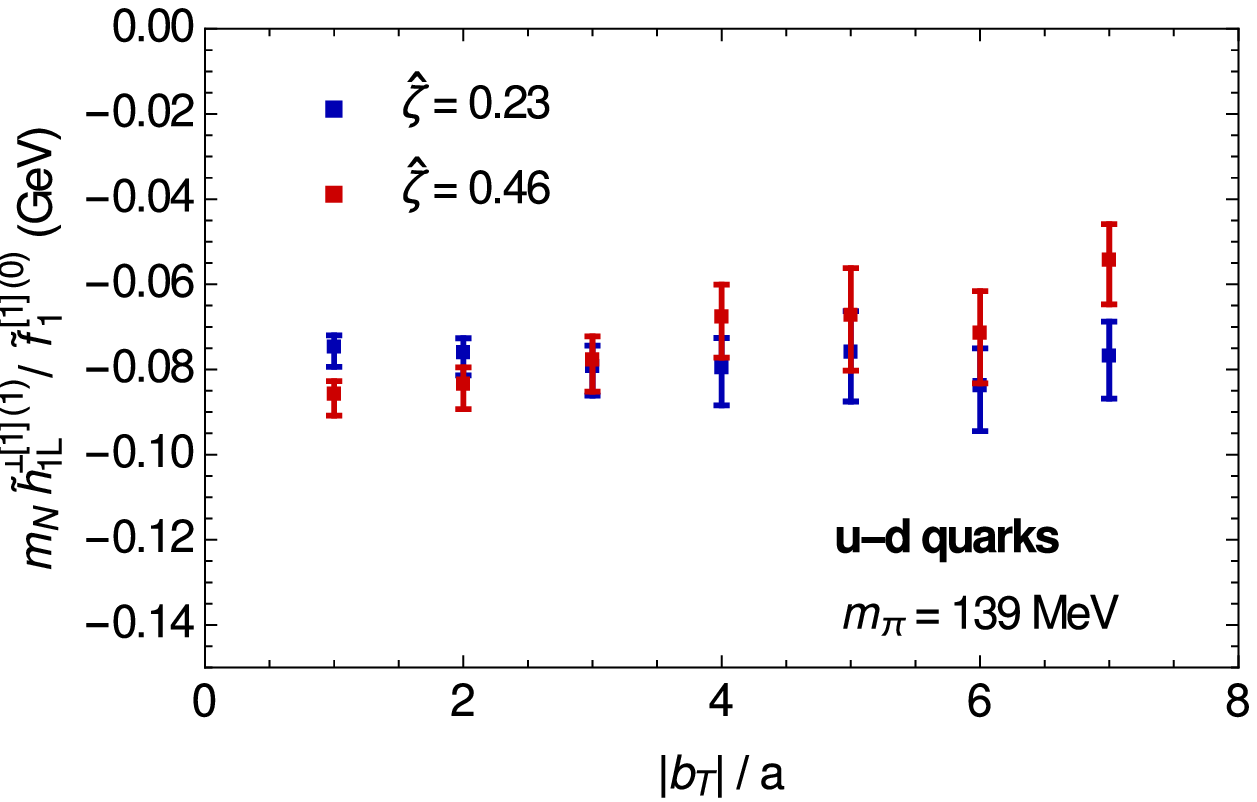}
\end{center}
\caption{Isovector generalized $h_{1L}^{\perp } $ worm-gear shift in the
nucleon. Left: Dependence on the staple length $\eta $ for fixed $|b_T |$
and two values of $\hat{\zeta } $; the extrapolation to the
$|\eta | \rightarrow \infty $ SIDIS/DY limit is described in the main
text. Right: SIDIS/DY limit as a function of $|b_T |$ for two values
of $\hat{\zeta } $. Error bars show statistical uncertainties only.}
\label{figetah}
\end{figure}

\begin{figure}
\begin{center}
\includegraphics[width=7.3cm]{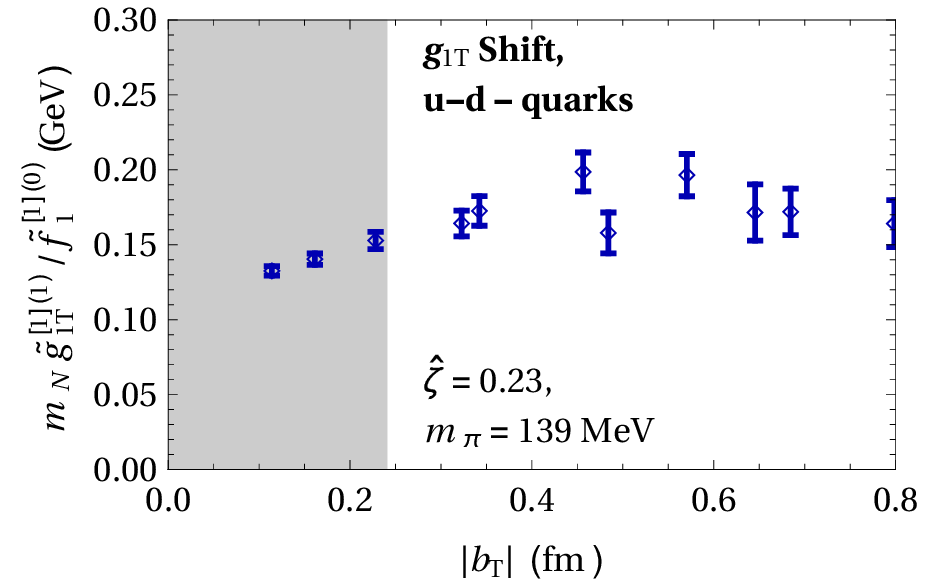}
\end{center}
\caption{
Isovector generalized $g_{1T} $ worm-gear shift in the nucleon in the
SIDIS/DY limit as a function of $|b_T |$ for $\hat{\zeta } =0.23$,
to be contrasted with Fig.~\ref{figetah} (right). Data in the shaded
region at small $|b_T |$ may be affected by significant finite lattice
spacing effects. Error bars show statistical uncertainties only.}
\label{figbg1T}
\end{figure}

Fig.~\ref{figetah} shows results for the isovector generalized 
$h_{1L}^{\perp } $ worm-gear shift in the nucleon. The left-hand panel
displays the dependence on the staple length $\eta $ for a fixed operator
separation $|b_T |$; data for two values of $\hat{\zeta } $ are shown.
In this observable, appreciable final/initial state interaction effects
are seen as one increases $|\eta |$ to approach the SIDIS/DY
limits. This worm-gear shift is again a T-even observable, and the
extrapolation $|\eta | \rightarrow \infty $ is obtained by averaging
over the plateau fits indicated by the horizontal lines. As with the
generalized axial charge, the data for the two values of $\hat{\zeta } $
do not differ significantly, suggesting a mild dependence on $\hat{\zeta } $.
The right-hand panel summarizes the results for the
$|\eta | \rightarrow \infty $ SIDIS/DY limit as a function of $|b_T |$.
Again, the dependence on $|b_T |$ is mild. The blue $\hat{\zeta } =0.23$
data are not distinguishable from a constant within the statistical
uncertainties and the red $\hat{\zeta } =0.46$ data exhibit at most
a weak trend as $|b_T |$ is varied. Interesting, however, is the
magnitude of the result: A very robust feature of a wide variety of
quark models that differ from one another qualitatively in many other
respects is that the $h_{1L}^{\perp } $ and $g_{1T} $ worm-gear TMDs are
predicted to have the same magnitude (and opposite sign). This includes
the spectator model \cite{spect}, the light-front constituent quark
model \cite{lfcqm}, the bag model \cite{bagm}, the light-front
quark-diquark model \cite{lfdiq}, the light-front version of the chiral
quark-soliton model \cite{solit}, and the covariant parton model \cite{covpm},
in which also the nonrelativistic limit can be consistently formulated.
On the other hand, Fig.~\ref{figbg1T} displays results for the $g_{1T} $
worm-gear shift obtained in a transversely polarized nucleon in a lattice
calculation on the same ensemble with the same computational methodology
(including the source-sink separation $8a=0.91\, \mbox{fm} $). Comparing
with the right-hand panel in Fig.~\ref{figetah}, the $h_{1L}^{\perp } $
and $g_{1T} $ worm-gear shifts do have opposite sign, but the obtained
magnitudes differ significantly, by close to a factor 2. Even though,
as has been emphasized above, the present computational setup is still
beset by significant systematic uncertainties, the discrepancy between the
magnitudes of the $h_{1L}^{\perp } $ and $g_{1T} $ worm-gear shifts seen
in Figs.~\ref{figetah} and \ref{figbg1T} appears sufficiently marked as
to make an explanation purely through systematic biases in the calculational
scheme seem unlikely. The discrepancy rather appears more likely to be a
manifestation of strong gluonic dynamical effects present in full QCD
that are not captured by quark models.

\section{Summary}
As part of an ongoing program exploring TMD observables within Lattice QCD,
the present investigation focused on longitudinally polarized nucleons.
Results were obtained for the generalized axial charge (\ref{genac}) and
the generalized $h_{1L}^{\perp } $ worm-gear shift (\ref{wgshift}), which
had hitherto not been studied; all leading-twist quark TMD observables have
thereby now been explored within the lattice TMD program. The data were
generated using a RBC/UKQCD domain wall fermion ensemble at the pion mass
$m_{\pi } = 139\, \mbox{MeV} $. To mitigate statistical fluctuations, a
fairly small source-sink separation $0.91\, \mbox{fm} $ was employed in
this exploratory calculation, leading to appreciable excited state
contaminations that are manifested in the magnitude of the extracted
generalized axial charge. The most striking outcome of this study is
the magnitude found for the generalized $h_{1L}^{\perp } $ worm-gear shift,
which differs by close to a factor 2 from its counterpart, the $g_{1T} $
worm-gear shift associated with longitudinally polarized quarks in a
transversely polarized nucleon. In a wide range of quark models, these
two worm-gear shifts are predicted to have the same magnitude (and
opposite sign, as is indeed found within the present investigation).
Despite the significant systematic biases still contained in the
calculational scheme employed, a discrepancy this marked appears
likely to reflect genuine physical effects generated by the gluonic
dynamics in the nucleon that are not captured by quark models.
Additional studies comparing the two worm-gear shifts with reduced
systematic uncertainties are necessary to draw more definite conclusions
regarding this interpretation.

\acknowledgments
Computing time granted by the John von Neumann Institute for
Computing (NIC) and provided on the supercomputer JURECA \cite{jureca}
(Booster module) at J\"ulich Supercomputing Centre (JSC) is gratefully
acknowledged, as are resources provided by the U.S.~DOE Office of Science
through the National Energy Research Scientific Computing Center (NERSC),
a DOE Office of Science User Facility, under Contract No.~DE-AC02-05CH11231.
Calculations were performed employing the Qlua \cite{qlua} software suite.
S.M.~is supported by the U.S.~DOE, Office of Science, Office of High
Energy Physics under Award Number DE-SC0009913.
M.E., J.N.~and A.P.~are supported by the U.S.~DOE, Office
of Science, Office of Nuclear Physics through grants numbered
DE-FG02-96ER40965, DE-SC-0011090 and DE-SC0023116,
respectively. S.S.~is supported by the National Science Foundation under
CAREER Award PHY-1847893. This work was furthermore supported by the
U.S.~DOE through the TMD Topical Collaboration.

\end{document}